\documentclass[fleqn,usenatbib]{mnras}

\usepackage[T1]{fontenc}

\DeclareRobustCommand{\VAN}[3]{#2}
\let\VANthebibliography\thebibliography
\def\thebibliography{\DeclareRobustCommand{\VAN}[3]{##3}\VANthebibliography}

\usepackage{graphicx}	% Including figure files
\usepackage{amsmath}	% Advanced maths commands
\usepackage{amssymb}	% Extra maths symbols
\usepackage{xcolor}
\usepackage{nicefrac}

\usepackage{comment}
\usepackage{newtxtext,newtxmath}

\DeclareRobustCommand{\mc}[1]{{\begingroup\color}}

\providecommand{\figfolder}{./figures/}
\title[Radiation from dual jet interactions in SMBHBs]{Non-thermal radiation from dual jet interactions in supermassive black hole binaries}

\author[E. M. Guti\'errez et al.]{
Eduardo M. Guti\'errez,$^{1,2,3}$\thanks{E-mail: emgutierrez@psu.edu}
Luciano Combi,$^{1,4,5}$ \thanks{{CITA National Fellow}}
Gustavo E. Romero$^{1,6}$
and Manuela Campanelli$^{7}$
\\
$^{1}$Instituto Argentino de Radioastronom\'ia (IAR, CCT La Plata, CONICET/CIC), C.C.5, (1984) Villa Elisa, Buenos Aires, Argentina\\
$^{2}$Institute for Gravitation and the Cosmos, The Pennsylvania State University, University Park PA 16802, USA\\
$^{3}$Department of Physics, The Pennsylvania State University, University Park PA 16802, USA\\
$^{4}$Perimeter Institute for Theoretical Physics, Waterloo, Ontario N2L 2Y5, Canada\\
$^{5}$Department of Physics, University of Guelph, Guelph, Ontario N1G 2W1, Canada \\
$^{6}$Facultad de Ciencias Astron\'omicas y Geof\'isicas, Universidad Nacional de La Plata, Paseo del Bosque s/n, 1900 La Plata, Buenos Aires, Argentina \\
$^{7}$Center for Computational Relativity and Gravitation, Rochester Institute of Technology, Rochester, NY 14623, USA
}

\date{Accepted XXX. Received YYY; in original form ZZZ}

\pubyear{2022}

\begin{document}
\label{firstpage}
\pagerange{\pageref{firstpage}--\pageref{lastpage}}
\maketitle

\begin{abstract}

    Supermassive black hole binaries (SMBHBs) are natural by-products of galaxy mergers and are expected to be powerful multi-messenger sources.
    They can be powered by the accretion of matter and then radiate across the electromagnetic spectrum, much like normal active galactic nuclei (AGNs).
    Current electromagnetic observatories have a good chance of detecting and identifying these systems in the near future. However, precise observational indicators are needed to distinguish individual AGNs from SMBHBs.
    In this paper, we propose a novel electromagnetic signature from SMBHBs: non-thermal emission produced by the interaction between the jets ejected by the black holes.
    We study close SMBHBs, which accrete matter from a circumbinary disc and the mini-discs formed around each hole.
    Each black hole ejects a magnetically dominated jet in the direction of its spin through the Blandford--Znajeck mechanism.
    We argue that in such a situation, the interaction between the jets can trigger strong magnetic reconnection events, where particles are accelerated and emit non-thermal radiation.
    Depending on whether the jets are aligned or misaligned, this radiation can have different periodicities.
    We model the evolution of the particles accelerated during the dual jet interaction and calculate their radiative output, obtaining spectra and providing estimates for the variability timescales.
    We finally discuss how this emission compares with that of normal AGNs.
    \end{abstract}


%%%%%%%%%%%%%%%%%%%%%%%%%%%%%%%%%%%%%%%%%%%%%%%%%%
%%%%%%%%%%%%%%%%% BODY OF PAPER %%%%%%%%%%%%%%%%%%

\section{Introduction} \label{sec:intro}

    Most luminous galaxies harbour a supermassive black hole (SMBH) at their centre \citep{kormendy1995inward, magorrian1998demography}.
    Since galaxies grow through sequential mergers of smaller galaxies, a potential outcome of such a process is the formation of SMBH binaries (SMBHBs) \citep{KormendyHo2013}.
    Understanding how these binaries are formed and how to find them is crucial for various areas of astrophysics, from cosmology to black hole accretion.
    
    After the coalescence of two galaxies, stellar dynamical friction and/or dissipation by surrounding gas leads the SMBHs to sink in the galactic remnant, forming a `hard' SMBH binary (SMBHB) and reducing their separation from $\sim$ kpc to $\sim$ pc scales \citep{Merritt2005}.
    At $\sim$ pc separations, dynamical friction ceases to be effective, and three-body stellar scattering \citep{Begelman_etal1980,quinlan1996dynamical} becomes the main mechanism to remove angular momentum from the binary.
    Recent galaxy formation studies show that the bulge of galaxies tends to have an asymmetric shape, ensuring a continuous refilling of stars available for scattering \citep{Vasiliev_etal2015} and avoiding the so-called `final parsec' problem \citep{milosavljevic2003final}, where the binary separation would stall at pc scales.
    Finally, if the SMBHs reach $\sim$ milli-parsec separations, the binary evolution becomes dominated by gravitational radiation and the black holes eventually merge in less than a Hubble time \citep{Peters:1964zz}. 
    
    SMBHBs at sub-parsec separations would be powerful gravitational wave (GW) emitters \citep{burke2019astrophysics}.
    The frequency band of these GWs ranges from nanohertz, typical of the inspiral phase of most massive binaries, to millihertz, typical of the proper merger (chirp).
    Recently, the pulsar timing array NANOGrav collaboration \citep{arzoumanian2020nanograv} as well as other collaborations \citep{reardon2023search, antoniadis2023second,lee2023searching} have provided strong evidence of the existence of a stochastic low-frequency GW background, likely coming from the superposition of unresolved SMBHB sources \citep{NANOGrav15yr_2023a, NANOGrav15yr_2023b}.
    A direct detection of individual SMBHB systems through GWs, however, will have to wait until the planned Laser Interferometer Space Antenna (LISA) mission is on \citep{mangiagli2022massive,engel2022advancing}, or the capabilities of Pulsar Timing Arrays (PTAs) are greatly improved. 
    Our best chance to find an SMBHB system in the near future is through the detection of its electromagnetic emission \citep{bogdanovic2021electromagnetic}.

    Identifying the presence of an SMBHB in an AGN is a difficult task.
    Most of the power from accretion discs in AGNs comes out very close to the central black hole.
    These scales are expected to be, overall, much shorter than the binary separation, and so the luminosity of an accreting SMBHB should differ only by a small fraction from that of a normal AGN.
    Up to this date, there has been no confirmed identification of a sub-parsec SMBHB source despite a growing number of candidates (e.g. \citealt{valtonen2008massive, Dotti2009,jiang2022tick}).
    
    Since we cannot resolve the deep interior of most galaxies, we need other robust signatures to find these binary sources.
    Some of the proposals include looking for periodic signals in the light curves \citep{Romero2000A&A,Romero2003,Valtonen2006, Graham2015a, Graham2015b, Liu2019, Saade2020} as well as Doppler variations \citep{DOrazio15}, broad emission line shifts \citep{Torres2003ApJ, Dotti2009}, a distinctive “notch” feature in the optical/IR spectrum \citep{Sesana_etal2011, Roedig:2014}, accretion-driven periodicities in the thermal spectrum of mini-discs  \citep{Bowen2019, Combi2021b, armengol2021circumbinary, Gutierrez_etal2022}, and X-ray periodicities \citep{Sesana_etal2011, Roedig:2014, Gutierrez_etal2022}.
    
   The accretion flow onto an SMBHB is expected to be qualitatively different from a standard disc surrounding a single BH.
   In particular, if the accreted matter has enough angular momentum and the mass ratio of the system is not too small, $q:=m_2/m_1 \gtrsim 0.1$, the torques exerted by the binary on the ambient gas carve a cavity of size $\sim 2 r_{12}$, where $r_{12}$ is the semi-major axis of the orbit, surrounded by a circumbinary disc (CBD) \citep{MM08}. 
   Several 2D hydrodynamical \citep{Farris11, DOrazio13, dittmann2021, Munoz2020a, Munoz2019, dittmann2023decoupling} and 3D magnetohydrodynamical simulations \citep{noble2012x,noble2021, armengol2021circumbinary, paschalidis2021minidisk, Gold_2014a, Combi2021b}, both in Newtonian gravity and full General Relativity, have shown the robust formation of such a cavity.
   Moreover, for cold discs, a strong $m=1$ overdensity mode, known as the `lump', forms in the inner boundary of the cavity.

   Simulations have also shown that accretion onto the binary is not mediated by viscous stresses but by thin streams that free-fall from the inner edge of the disc to the black holes with relatively low angular specific angular momentum $\ell$.
   If $\ell$ is larger than the specific angular momentum at the innermost stable circular orbit (ISCO) radius, the matter will start orbiting around the black hole forming a mini-disc \citep{Miller:2013ApJ, Tanaka12, Bowen2017}.
   Turbulent stresses in the mini-disc will transport angular momentum outward and the mini-disc will expand until it reaches the truncation radius (or Hill's sphere), $r_{\rm t}$, where the disc is disrupted by the tidal forces of the other BH.

   For typical AGNs, both the CBD and the mini-discs are expected to radiate a thermal spectrum. 
   The global spectrum of the system may be, however, different from that of a single BH disc. 
   First, because of the presence of the circumbinary cavity, the spectrum would a priori show a gap between the thermal peaks of the CBD and mini-discs \citep{Roedig:2014} (although it is unclear from simulations how gas-depleted the cavity is);  moreover, for small separations, $\ell$ can be, on average, smaller than the ISCO, reducing the radiative efficiency of the mini-discs \citep{Gutierrez_etal2022}.

    If the binary contains rotating black holes, the magnetic fields carried by the mini-discs can fill the ergosphere of each black hole and a pair of jets can be launched via the Blandford--Znajeck (BZ) mechanism \citep{BlandfordZnajek1977}; this has been demonstrated by force-free simulations \cite{moesta2012detectability, palenzuela2010dual} as well as GRMHD simulations \cite{kelley2021gravitational, paschalidis2021minidisk, Combi2021b, giacomazzo2012, cattorini2021}. 
    These simulations have shown that at the merger, the Poynting luminosity increases up to an order of magnitude, which could be translated to an increased EM emission, though it is probably outshined by the galaxy luminosity if most of the emission is produced in the radio band \citep{kaplan2011blindly}.

    The emergence of a dual jet structure in SMBHBs might give rise to characteristic EM signatures different from those of single BH jets.
    For example, assuming the emitted luminosity is tied to the Poynting luminosity of the system, MHD variabilities of the accretion rate could be imprinted in the emission as well \citep{Combi2021b,Paschalidis_etal2021, bright2022minidisk} (see Sec. \ref{sec:variab}). 
    In this work, we propose a novel mechanism where non-thermal radiation is produced due to the interaction between jets launched by SMBHBs in the inspiralling regime.
    More precisely, we argue that the region of contact between the jets is prone to the formation of large-scale reconnection layers where magnetic energy is released and a fraction of the particles in the jet are accelerated up to high-energies and radiate through synchrotron and inverse Compton processes.
    We build a simple semi-analytical model to estimate the electromagnetic outcome from such interactions.

    The paper is structured as follows.
    In Section \ref{sec:physical_scenario}, we describe the physical scenario under consideration and introduce the main characteristics of the model.
    The model includes the calculation of the electromagnetic radiation from the CBD and the mini-discs, plus a detailed treatment on how jets interact and how particles are accelerated and radiate at the reconnection layer that forms between the jets.
    In Section \ref{sec:results}, we show the results of the calculation of spectral energy distributions (SEDs), exploring its dependence on the orbital parameters as well as on less constrained microphysical properties of the plasma.
    We also discuss the potential periodicities associated to this emission, differentiating the cases where the two jets are aligned or misaligned.
    In Section \ref{sec:discussion}, we discuss the prospects of detecting the proposed phenomena, comparing its main features with those of standard jetted AGNs.
    Finally, in Section \ref{sec:conclusions}, we present the conclusions of our research.

\section{Physical model}
\label{sec:physical_scenario}
   
\begin{figure}
    \centering
    \includegraphics[width=0.999\linewidth]{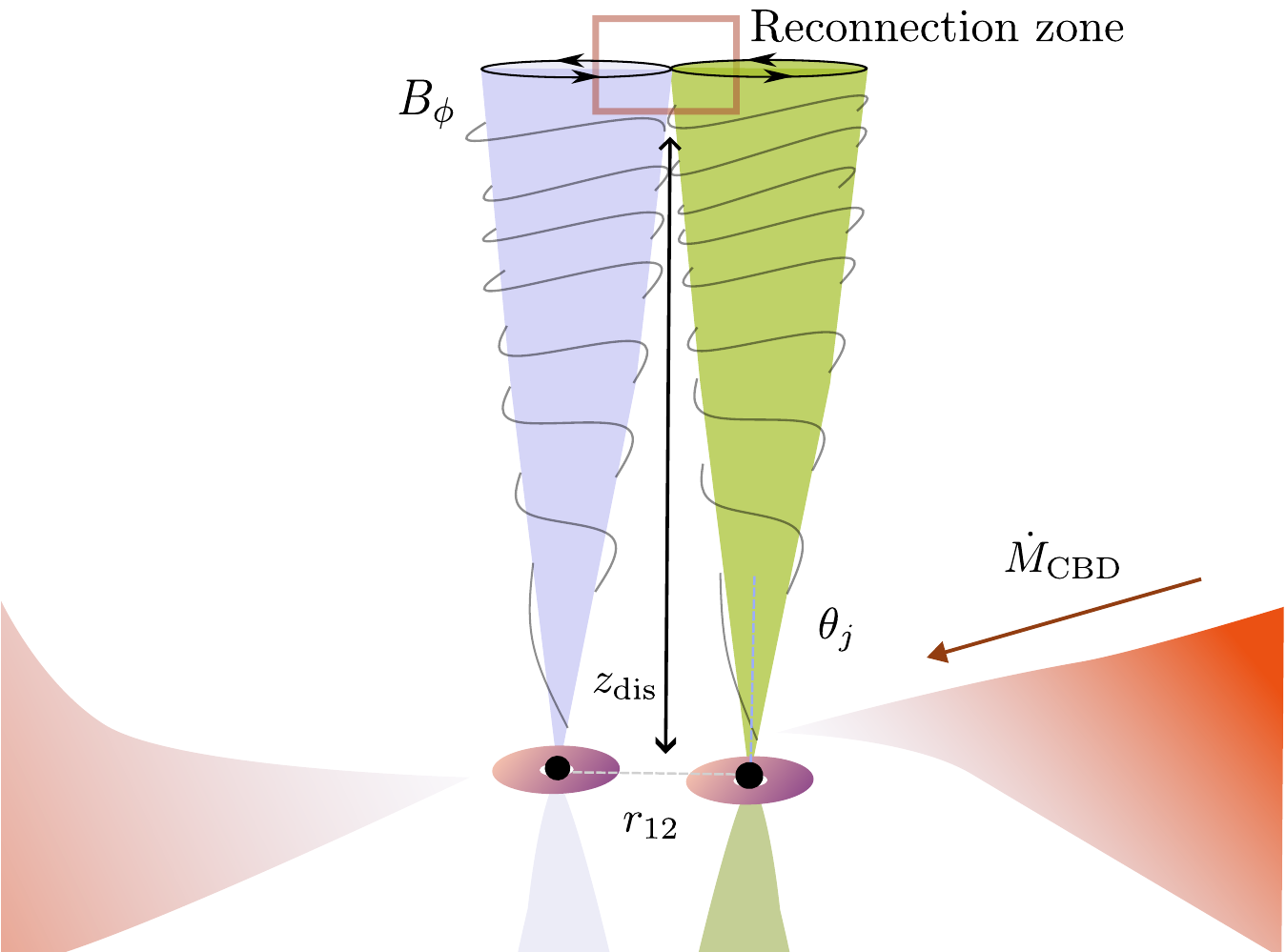}
    \caption{
    Schematic diagram of an accreting SMBHB with jets.
    In the image, the mini-discs, jets, and CBD are shown (as well as the streams through which the mini-discs are fed).
   The toroidal components of the magnetic field in both jets have opposite polarities and give rise to the formation of a `reconnection layer' between the jets at a height $\sim z_{\rm dis}$.}
    \label{fig:smbhb_scheme}
\end{figure}

    The physical scenario considered is shown schematically in Figure \ref{fig:smbhb_scheme}.
    An equal-mass SMBHB system at close separations ($r \lesssim 200 R_{\rm g}$) is surrounded by an accretion flow, whose structure consists of a CBD and mini-discs around each black hole.
    The two black holes are rapidly spinning and the gas falling from the CBD carries enough magnetic flux for the Blandford--Znajek mechanism to work efficiently and launch a pair of Poynting-flux dominated jets.
    At some height above the orbital plane, the magnetically-dominated jets get close enough to each other and interact, dissipating magnetic energy.
    Such a dissipation may occur if the magnetic field carried by the jets, in particular its toroidal component, has partially opposite polarities at the interaction region. Then, at the contact surface between the jets, a large-scale reconnection layer will form. The jet interaction might also excite other magnetic instabilities similar to the interaction of (non-relativistic) flux tubes; however, there are yet no ideal MHD simulations exploring this scenario.
   
    A fraction of the magnetic energy released in the reconnection region is transferred to the particles in the jet which are accelerated to a non-thermal energy distribution; these particles, then, cool down by interacting with the magnetic and radiation fields present in the region through synchrotron and synchrotron-self Compton (SSC).
    
    In what follows, we develop a simple model to characterise the jet-jet interaction envisioned above and estimate the electromagnetic emission produced.
    For completeness, we also include in our treatment the thermal emission from the CBD and the mini-discs, since they are linked to the jets and its emission will likely be present too.
    
    \textbf{Conventions:} Non-primed kinematic quantities $(p)$ refer to the laboratory frame, single-primed quantitites $(p')$ refer to the fluid frame. %We use cgs units.
    
\subsection{Binary black hole spacetime}
\label{sec:orbital_motion}

    Let us consider two supermassive black holes with masses $m_i$ ($i=1,2$), and normalised dimensionless spins $\chi_i$ forming a binary system.
    If the system has reached a stage where the emission of GWs is efficient and dominates the evolution of the orbit ($r_{12} \lesssim 10^3 R_{\rm g}$), it will slowly evolve through quasi-circular orbits of decreasing radius.
At 1st post-Newtonian order, the coalescence time can be estimated as \citep{Peters:1964zz}
\begin{equation}
    t_{\rm c} = \frac{5}{256} \frac{c^5}{G^3} \frac{r_{12}(t_0)^4}{M^2 \mu} \sim 36~ \frac{(1+q)^2}{q} \left( \frac{r_{12}}{30 R_{\rm g}} \right)^4 M_7 ~{\rm days},
    \label{eq:t_coal}
\end{equation}
$\mu := qM / (1+q)^2$ is the reduced mass of the system, and $M = 10^7 M_7~ {\rm M}_\odot$.
    Under this approximation, the orbital frequency of the binary system is given by the Keplerian value, and the orbital period is
\begin{equation}
	P_{\rm bin} \simeq 14 ~ M_7 \left( \frac{r_{12}}{30R_{\rm g}} \right)^{3/2}~{\rm hr}.
 \label{eq:Pbin}
\end{equation}
	In what follows, we assume that the SMBHB system has a mass ratio $q=m_1/m_2 \approx 1$ and is evolving in quasi-circular orbits.

\subsection{Circumbinary and mini discs accretion: thermal emission}
\label{sec:discos}

    The SMBHB system is surrounded by a CBD, which is known to have an inner edge at $\sim 2r_{12}$ \citep{MunozLai16,ArtymLubow94,Noble12} produced by binary torques counteracting the viscous stresses in the disc.
	The gas falls from the inner edge of the CBD onto the binary, and part of this gas has enough angular momentum to form mini-discs around each BH \citep{Miller:2013ApJ,Roedig:2014, Tanaka12}. 
	The mini-discs extend from the tidal truncation radius \citep{Paczynski:1977, Roedig:2014}, $r_{\rm t} \sim 0.3 q^{-0.3} r_{12}$ $(0.3 q^{0.3} r_{12})$ for the primary (secondary), down to the radius of the inner-most stable orbit (ISCO), $r_{\rm isco} \gtrsim R_{\rm h}$ for rapidly spinning BHs, where $R_{\rm h} \sim R_{\rm g}$ is the event horizon radius.
 
   We parameterise the accretion rate in the mini-disc around the $i$th BH as $\dot{M}_{i} = f_{i} \dot{M}_{\rm cbd}$, where $f_{1} + f_{2} \lesssim 1$ and $\dot{M}_{\rm cbd}$ is the steady-state accretion rate in the CBD.
    If the binary inspiral speed, $\dot{r}_{12} \sim r_{12}^{-3}$, becomes faster than the radial velocities triggered by the accretion process, $\dot{r}_{\rm disc} \sim -3\nu/2r_{\rm edge}$, with $\nu$ the kinematic viscosity coefficient, the binary (and mini-discs) will decouple from the CBD and the accretion will be out of inflow equilibrium, $f_{1,2} \ll 1$ \citep{dittmann2023decoupling, major2023disappearing}.
   Simulations also show that the mini-disc accretion rate is variable, modulated by the binary frequency \citep{Bowen2019, Combi2021b, westernacher2021, Munoz2016}.
   We focus here on the time-averaged properties of the accretion and leave the discussion on the potential impact of variability in our model to Section \ref{sec:variab}.

    We are interested here in bright AGN-like discs which can power jets.  At accretion rates close to Eddington, the CBD can be modelled as a geometrically-thin Shakura--Sunyaev disc (SSD, \citealt{ShakuraSunyaev1973}) with its inner boundary at $r_{\rm edge} \approx 2r_{12}(t)$ around an accreting object of mass $M=m_1+m_2$, emitting as a multi-temperature black body (see, e.g., \citealt{DermerMenon2009}).
    This approximation provides a good agreement with the spectrum that results from the more detailed ray-tracing of GRMHD data \citep{Gutierrez_etal2022}.
    Mini-discs, however, are rather different from single black hole accretion discs.
    In particular, at short binary separations, a significant part of the matter that falls from the inner edge of the CBD has low specific angular momentum and it free-falls to the holes without dissipating much energy \citep{Combi2021b, Gutierrez_etal2022}.
    Moreover, mini-discs may reach higher temperatures than those expected in single-BH systems \citep{dAscoli2018, Gutierrez_etal2022} due to streams shocking and heating the disc.

    To account for these two effects in a simple way, we introduce two minimal modifications to the SSD solution.
\begin{enumerate}
	\item We assume that only a fraction $f_{\rm d}<1$ of the matter in the mini-discs has enough angular momentum to form a disc, and we use an ``effective accretion rate'' as $\dot{M}_{{\rm eff},i} = f_{\rm d} \dot{M}_i = f_{\rm d} f_i \dot{M}_{\rm cbd}$.
	\item We include a colour factor $f_{\rm col}=2$ \citep{shimura1995}, so that the effective temperature and the spectral flux are modified as $T_{\rm eff}(r) \rightarrow f_{\rm col} T_{\rm eff}(r)$ and
	$F_\nu \rightarrow f_{\rm col}^{-4} F_\nu$, respectively, conserving the bolometric luminosity.
\end{enumerate}

    In summary, we model the spectrum of each mini-disc as that of a Shakura--Sunyaev disc around a black hole of mass $m_i$ and with accretion rate $\dot{M}_{{\rm eff},i}$, extending from $r_{{\rm isco},i}$ to the truncation radius $r_{{\rm t},i}$, plus the addition of a colour factor $f_{\rm col}$ to account for the higher temperatures achieved.

\subsection{Dual jets}
\label{sec:jets}

    If enough magnetic flux is carried by matter falling onto the cavity, the mini-discs can launch powerful jets via the BZ mechanism.
    Considering the well-known `disc-jet connection' \citep{FalckeBiermann1995}, we assume that the power of each jet is proportional to the total accretion power onto the corresponding black hole:
\begin{equation}
    L_{{\rm j},i} = \frac{1}{2}\eta \dot{M}_i c^2 \simeq 6.3 \times 10^{44} \eta_{-1} \left( \frac{f_i}{0.5} \right) \dot{m} M_7 ~~~{\rm erg~s}^{-1},
    \label{eq:Lj}
\end{equation}
where $\eta= 10^{-1} \eta_{-1}$ is the jet launching efficiency, $\dot{M}_{\rm cbd} = \dot{m} \dot{M}_{\rm Edd}$, with $\dot{M}_{\rm Edd}:=1.39 \times 10^{18} (M/{\rm M}_\odot)$ the Eddington accretion rate for the CBD, and the factor $1/2$ corresponds to a one-sided jet.
    Recent GRMHD simulations of SMBHB's mini-discs \citep{Paschalidis_etal2021, Combi2021b} suggest that $\eta \lesssim 0.1$, though this value could be higher for more favourable conditions as those of magnetically arrested discs \citep{tchekhovskoy2011}.
    
    For Poynting-flux dominated jets to radiate their energy away, magnetic energy must be converted to internal energy of the plasma. 
    In AGN jets, dissipation of magnetic energy can occur efficiently through reconnection within compact regions of the jet.
    This, however, tends to occur at large distances from the black holes \citep{Giannios2013, nalewajko2012energetic}.
    As we argue below, jetted SMBHBs may dissipate part of their magnetic energy very close to the black holes, different from single BH jets (see Sec. \ref{sec:discussion}).

    At close binary separations, the two jets can easily interact if they have non-zero opening angles or if they are misaligned.
    Provided that no significant bending or twisting of the jets occurs, the height at which the two jets interact can be estimated as $z_{\rm dis} \sim (r_{12}/2)/\theta_*$, where $\theta_*$ parameterises our ignorance about the inclination angle of the flow lines in the jet.
    In the case of aligned jets, we may expect $\theta_* \approx \theta_{\rm j}$, though it can be higher for inclined jets.
    
    For a highly magnetised jet, the magnetic energy density in the jet at the dissipation height is $u'_B \approx L_{{\rm j},i}/(2\pi \theta_{\rm j}^2z_{\rm dis}^2 c \Gamma_{\rm j}^2)$; then, if the magnetic field is predominantly toroidal, its intensity is given by $B'\simeq 2\sqrt{L_{{\rm j},i}/c}/(\Gamma_{\rm j}z_{\rm dis}\theta_{\rm j}) \simeq 4 L_{{\rm j},i}^{1/2} c^{-1/2} (\Gamma_{\rm j} r_{12})^{-1} (\theta_*/\theta_{\rm j})$.
    We can recast this in the following useful form:
\begin{multline}
  B' \approx 8.3 \times 10^3 \eta_{-1}^{1/2} \dot{m}^{1/2} M_7^{1/2} \times \\
  \left( \frac{f_i}{0.5}\right)^{1/2}\left(\frac{\Gamma_{\rm j}}{1.5}\right)^{-1} \left( \frac{r_{12}}{30 R_{\rm g}} \right)^{-1} \left( \frac{\theta_*}{\theta_{\rm j}}\right)~{\rm G}.
  \label{eq:B}
\end{multline}
	The expression above gives the magnetic field intensity at the interacting region and depends solely on the global properties of the binary and the accretion flow.
 In particular, for aligned spins, $\theta_* \approx \theta_{\rm j}$ and the magnetic field is independent of the jet opening angle.

  The interaction of the jets at a height $z_{\rm dis}$ in our model resembles the collision of (relativistic) magnetised flux tubes.
  This could give rise to one or both of the following effects: {\it a)} a twisting and bending of the jets due to the increasing magnetic pressure that each jet imprints in the other, and {\it b)} encounters of regions with magnetic fields of opposite polarities at high speed.
  The first situation would give rise to the development of kink instabilities, which are known to trigger magnetic reconnection when different portions of the twisted jet encounter \citep{Alves_etal2018, MedinaTorrejon_etal2021}.
  The second situation would directly give rise to a large-scale magnetic reconnection layer at the contact surface between the jets.
  Either way, magnetic energy will be efficiently released and transferred to the plasma in the form of bulk kinetic energy or internal energy.

    In our model, we assume that the interaction between the two jets leads to the formation of current sheets (either by the direct collision or as a by-product of the development of magnetic instabilities) and the occurrence of large-scale magnetic reconnection events where particles are accelerated.
    The typical length-scale for the reconnection layer, $l$, in our scenario is of the order of the cylindrical radius of the jets at the dissipation height, namely $l \sim \theta_{\rm j} z_{\rm dis} \sim r_{12}/2$.
    At these close distances to the holes, jets are not expected to be baryon loaded \citep{RomeroGutierrez2020,Romero2021AN}, so we will limit our model to pair-dominated jets.

    \subsubsection{Particle acceleration and evolution}
    \label{sec:particle}
    
    Numerous particle-in-cell (PIC) simulations show that magnetic reconnection in large-scale current sheets results in the formation of a series of magnetised plasmoids of different sizes \citep{Samtaney_etal2009, Uzdensky_etal2010, SironiSpitkovsky2014, Sironi_etal2016, Petropoulou_etal2016, Petropoulou_etal2018}.
    In the plasmoid frame, particles are accelerated and form an isotropic (non-thermal) power-law particle distribution \citep{ZenitaniHoshino2001, SironiSpitkovsky2011, SironiSpitkovsky2014, Petropoulou_etal2019}.
    The spectral index of the distribution, $p$, strongly depends on the value of the magnetisation parameter in the non-reconnected plasma, i.e. in the jet, being lower (harder spectra) for higher magnetisation values \citep{SironiSpitkovsky2014, Sironi_etal2016, Werner_etal2016, Guo_etal2014, Guo_etal2015, Guo_etal2016, Guo_etal2021, Petropoulou_etal2019}.

    Magnetic energy is released through magnetic reconnection at a rate $\beta_{\rm rec}=v_{\rm rec}/c$, where $v_{\rm rec}$ is the reconnection speed.
    For relativistic reconnection, $v_{\rm rec} \lesssim v_{\rm A}$, where $v_{\rm A} \sim \sqrt{\sigma/(\sigma+1)}c$ is the Alfv\'en speed and $\sigma$ is the plasma magnetisation.
    The power transferred to the non-thermal particles in the plasma is then $L_{\rm e}'\approx u'_B \pi l^2 \beta_{\rm rec} c$.

    For simplicity, we assume that a single spherical blob of size $R'_{\rm b} \sim l$ contains the accelerated particles, which are injected with a power-law spectrum in the blob's co-moving frame:
\begin{equation}
    Q'(\gamma') = Q_0' \gamma'^{-p} H[\gamma';\gamma'_{\rm min}, \gamma'_{\rm max}].
\end{equation}
    Here, $H[x;x_1,x_2]$ is the Heaviside function and $Q_0'$, $\gamma_{\rm min}'$, $\gamma_{\rm max}'$ are unknowns that we fix below.
    The normalisation $Q_0'$ is determined by the condition $m_{\rm e} c^2\int_{\gamma_{\rm min}'}^{\gamma_{\rm max}'}d\gamma' \gamma' Q'(\gamma') = L_{\rm e}'$.
    Simulations also show that only a fraction $f_{\rm e} \sim 0.3$ of the particles are accelerated \citep{Hoshino2022, French_etal2023}, so that the non-thermal particle injection rate is $\dot{N}_{\rm e}' = \int_{\gamma_{\rm min}'}^{\gamma_{\rm max}'}d\gamma' Q'(\gamma') \sim f_{\rm e} n_{\rm e}' \pi l^2 v_{\rm rec}$.
    Then, we can estimate the mean Lorentz factor as
\begin{equation}
    \langle \gamma' \rangle \sim L'_{\rm e} / (\dot{N}'_{\rm e} m_{\rm e}c^2)\sim 2 f_{\rm e}^{-1} \sigma',
    \label{eq:mean_gamma}
\end{equation}
since $u'_B/n_{\rm e}'= 2 \sigma' m_{\rm e} c^2$.

    Depending on whether $p>2$ (soft spectrum) or $p<2$ (hard spectrum), the mean Lorentz factor constrains the minimum and maximum Lorentz factor of the distribution, respectively.
    For a pair plasma with $p>2$, and assuming $\gamma_{\rm max}' \gg \gamma_{\rm min}'$, we can explicitly integrate the numerator and denominator in the expression of $\langle \gamma' \rangle$, and use Eq. \ref{eq:mean_gamma}, to get
\begin{equation}
    \gamma_{\rm min}' \approx 2 f_{\rm e}^{-1}\sigma' \frac{(p-2)}{(p- 1)}.
    \label{eq:gmin}
\end{equation}

    The maximum Lorentz factor in this case is determined by the balance between the acceleration and cooling timescales: $t_{\rm acc}'=t_{\rm cool}'$.
    The acceleration timescale can be parameterised in a general way as \citep{DermerMenon2009}
\begin{multline}
    t_{\rm acc}' = \frac{\gamma'}{\dot{\gamma}'} = \frac{ \gamma'm_{\rm e} c}{\zeta_{\rm acc} e B'}  \approx 5.7 \times 10^{-4}  \gamma' \zeta_{{\rm acc},-4}^{-1} B'^{-1} \simeq \\
     10^{-7}~ \gamma' (\Gamma_{\rm j}/1.5) \zeta_{\rm acc,-4}^{-1} {\eta_{-1}}^{-1/2} \dot{m}^{-1/2} M_7^{1/2}  \left( \frac{r_{12}}{30 R_{\rm g}} \right)~{\rm s},
    \label{eq:acc}
\end{multline}
where $e$ is the electron charge and $\zeta_{\rm acc}= 10^{-4} \zeta_{{\rm acc},-4}$ measures the acceleration efficiency, which may be as large as $\sim \beta_{\rm rec}$ in case of direct acceleration by the reconnection electric field $E_{\rm rec}'\sim \beta_{\rm rec} B'$, but much lower for 1st order Fermi acceleration (e.g., \citealt{delvalle2016}).
    The dominant cooling process for the non-thermal particles in our scenario is synchrotron radiation with a cooling timescale
\begin{multline}
    t_{\rm sy}'(\gamma') \approx 11 \times \left( \frac{B'}{4.5\times 10^3~{\rm G}} \right)^{-2} \gamma'^{-1}~{\rm s} \sim \\
    25 \gamma'^{-1}~ {\eta_{-1}}^{-1} \dot{m}^{-1} M_7 \left( \frac{\Gamma_{\rm j}}{1.5} \right)^2 \left( \frac{r_{12}}{30 R_{\rm g}} \right)^{2} ~{\rm s},
\end{multline}
    and thus the maximum Lorentz factor results
\begin{multline}
     \gamma_{\rm max}' = \sqrt{ \frac{6\pi e \zeta_{\rm acc}  }{ \sigma_{\rm T} B'} } \simeq \\
     1.26 \times 10^4 \zeta_{{\rm acc},-4}^{1/2} {\eta_{-1}}^{-1/4} \dot{m}^{-1/4} M_7^{1/4} \left( \frac{\Gamma_{\rm j}}{1.5} \right)^{1/2} \left( \frac{r_{12}}{30 R_{\rm g}} \right)^{1/2},
     \label{eq:gmax}
\end{multline}
where $\sigma_{\rm T}$ is the Thomson cross-section.

     On the other hand, if $p < 2$, most of the energy is carried by the highest energy particles, and $\gamma_{\rm max}'$ is directly limited by the magnetisation $\sigma'$ \citep{SironiSpitkovsky2014, Guo_etal2015, Werner_etal2016}:
\begin{equation}
     \gamma_{\rm max}' = \left[ 2\sigma' f_{\rm e}^{-1} \frac{(2-p)}{(p -1)} \right]^{1/(2-p)}.
     \label{eq:gmax_hard}
\end{equation}
    Here, $\gamma_{\rm min}'$ is not constrained and thus we simply set $\gamma_{\rm min}'=1$.  In addition to cooling down, particles will leave the blob with a timescale $t'_{\rm esc} \gtrsim R'_{\rm b}/c \sim 6.6\times 10^2 (r_{12}/30R_{\rm g}) M_7~{\rm s}$.
We then have in general $t_{\rm sy}' \ll t_{\rm esc}'$, namely we are always in the fast-cooling regime, and the particle energy distribution can be approximated as $N'(\gamma') \approx Q'(\gamma') t_{\rm sy}' \propto \gamma'^{-p+1}$.

\subsubsection{Non-thermal radiation}

    Once we have obtained the particle energy distribution in the comoving frame of the plasma, $N'(\gamma')$, we calculate the non-thermal radiation produced by synchrotron and SSC emission using the publicly available code \texttt{agnpy}\footnote{\url{https://github.com/cosimoNigro/agnpy}} \citep{Nigro2022}.
    The synchrotron and SSC emissivities, $j_{\nu'}'^{\rm (sy)}$ and $j_{\nu'}'^{\rm (ssc)}$, are calculated as outlined in \citet{Finke_etal2008}, and we consider the internal synchrotron-self-absorption (SSA) at lower energies and the absorption due to the photo-pair creation process ($\gamma \gamma \rightarrow {\rm e}^+ {\rm e}^-$) at high energies \citep{DermerMenon2009}.
    Finally, the observed spectral flux on Earth for a source at a luminosity distance $d_L$ (corresponding to a redshift $z$) is \citep{DermerMenon2009}
\begin{equation}
    \nu F_{\nu}= \left( \frac{3 u(\tau)}{\tau} \right) \frac{\mathcal{D}^4 V'_{\rm b}}{d_L^2} \nu' \left( j_{\nu'}'^{\rm (sy)} + j_{\nu'}'^{\rm (ssc)}\right),
\end{equation}
where $\nu = \mathcal{D} \nu'/(1+z)$, and
\begin{equation}
    u(\tau) = \frac{1}{2} + \frac{{\rm e}^{-\tau}}{\tau} - \frac{\left( 1 - {\rm e}^{-\tau} \right)} {\tau^2},
\end{equation}
where $\tau = 2 R'_{\rm b} (\kappa'_{\rm ssa} + \kappa'_{\gamma \gamma} )$ is the total internal optical depth, $\kappa'_{\rm ssa}$ and $\kappa'_{\gamma \gamma}$ are the absorption coefficients due to SSA and $\gamma \gamma$ absorption, respectively, $V_{\rm b}'=(4/3)\pi R_{\rm b}'^3$ is the blob's volume, and $\mathcal{D}$ is the Doppler factor of the blob.
The latter depends on the direction and the Lorentz factor with which the blob moves in the jet comoving frame.
    For simplicity, we assume that the blob is non-relativistic in the jet proper frame and thus it moves with the same bulk motion as the jet; hence, $
    \mathcal{D} = \left[ \Gamma_{\rm j} \left( 1 - \beta_{\rm j} \cos i \right) \right]^{-1}$, where $i$ is the viewing angle. In the next section, we show SEDs of observed luminosity defined as $L_\nu = 4\pi d_L^2 F_\nu$.

\section{Results}
\label{sec:results}

\begin{table}
 \caption{Parameters of the fiducial model.}
 \label{tab:anysymbols}
 \begin{tabular*}{\columnwidth}{@{}l@{\hspace*{20pt}}l@{\hspace*{30pt}}l@{}}
  \hline
  Parameter & Description & Fiducial Value\\
  \hline
  $M$ & Total mass of the system & $10^{7} M_\odot$ \\[2pt]
  $q$  & Mass ratio of the system & $1$\\[2pt]
  $r_{12}$  & Binary Separation & $30$ $R_{\rm g}$\\[2pt]
  $\chi_1, \chi_2$ & Norm. spins of the black holes & $0.9$\\[2pt]
  $i$ & Viewing Angle & $\arccos{\beta_{\rm j}}$\\[2pt]
  $\dot{M}_{\rm cbd}$  & Accretion rate of the CBD & $\dot{M}_{\rm Edd}$\\[2pt]
  $f_1, f_2$ & Mini-disc's accretion rate fraction & $0.5$ \\[2pt]
  $f_{\rm d}$ & Circularised acc. rate fraction & $0.2$ \\[2pt]
  $\eta$ & Jet launching efficiency & $0.1$ \\[2pt]
  $\Gamma_{\rm j}$ & Jet Lorentz factor & $1.5$ \\[2pt]
  $\sigma'$ & Jet comoving magnetisation & $10$\\[2pt]
  $p$ & Spectral Index & $2.1$\\[2pt]
  $f_{\rm e}$ & Fraction of particles accelerated & $0.3$\\[2pt]
  $\beta_{\rm rec}$ & Reconnection rate & $0.2$\\[2pt]
  $\zeta_{\rm acc}$ & Acceleration efficiency & $10^{-4}$, $0.2$\\[2pt]
  \hline
 \end{tabular*}
\end{table}

\begin{figure}
        \includegraphics[width=0.99\linewidth]{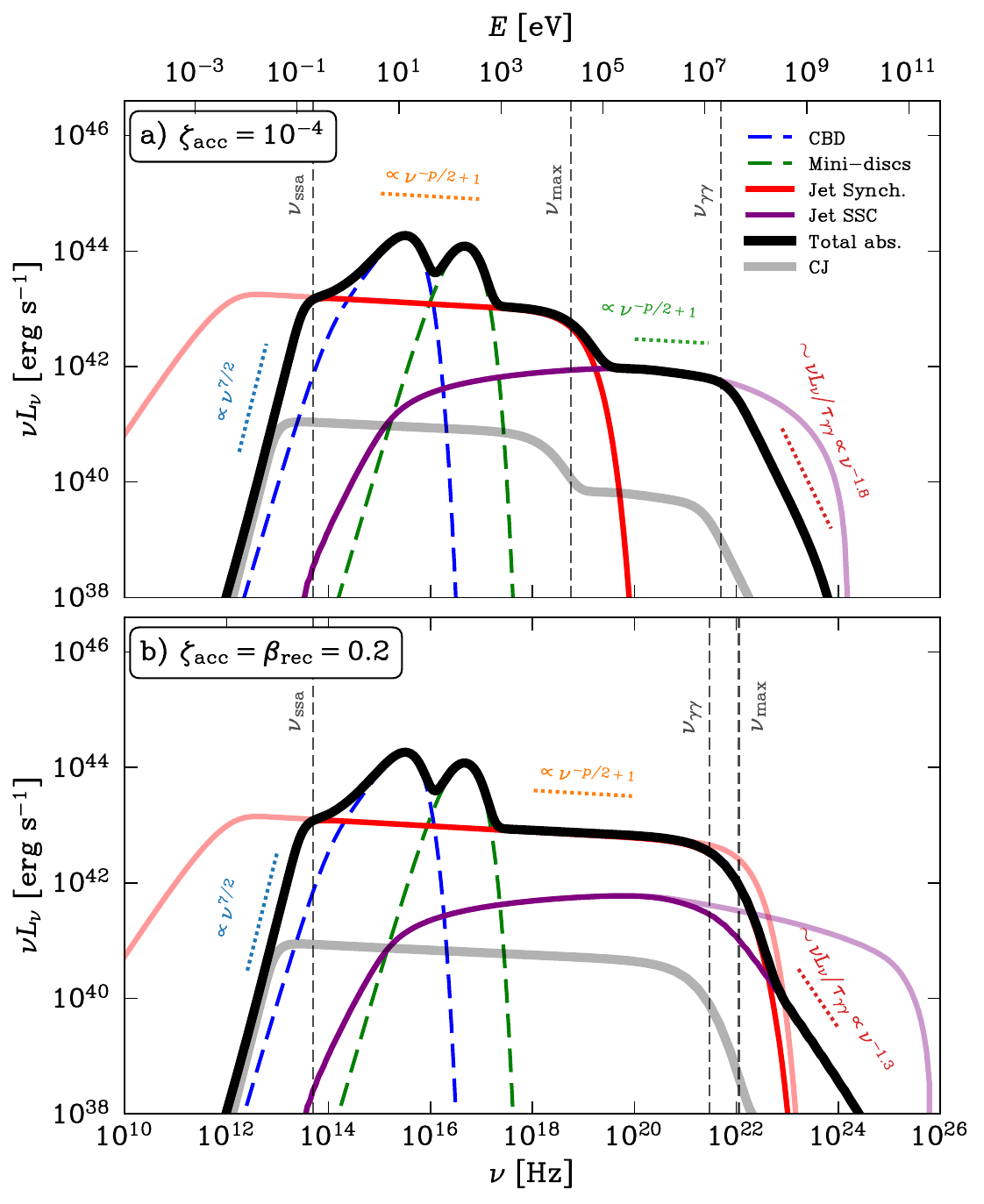}
        \caption{Spectral energy distribution for our reference model of an accreting SMBHB with a dual jet interaction occurring at $z_{\rm dis}$.
        The parameters of the model are given in Table \ref{tab:anysymbols}.
        {\bf a)} Low acceleration efficiency: $\zeta_{\rm acc}=10^{-4}$. {\bf b)} High acceleration efficiency: $\zeta_{\rm acc}=\beta_{\rm rec}=0.2$.
        The different contributions to the SED are shown in different line styles and colours: CBD (dashed blue line), mini-discs (dashed green line), dual jet synchrotron emission (solid red line), and dual jet SSC emission (solid purple line).
        The black solid line shows the total spectrum corrected by internal $\gamma \gamma$ absorption and SSA.
        The solid grey line shows the total spectrum of the counter-jet (CJ). The vertical dashed lines show the SSA frequency ($\nu_{\rm ssa}$), the maximum frequency for the synchrotron spectrum ($\nu_{\rm max})$, and the $\gamma \gamma$ absorption frequency, where $\tau_{\gamma \gamma}\sim 1$ ($\nu_{\gamma \gamma}$).
        We also show the spectral indices of the SED in each characteristic band.}
        \label{fig:SED_fiducial}
\end{figure}

    The physical model described in the previous section allows us to estimate the non-thermal emission produced by the interaction of dual jets in an SMBHB system at close separations.
    To explore this emission, we first define a reference model considering an SMBHB system with a total mass $M=10^7 M_\odot$ and a separation $r_{12}=30 R_{\rm g}$, surrounded by a CBD accreting at the Eddington rate, $\dot{M}_{\rm cbd}=\dot{M}_{\rm Edd}$.
    We also assume that the two black holes are identical ($q=1$), fast-rotating ($\chi_i \gtrsim 0.9$), and have mini-discs with $f_i=0.5$ and $f_{\rm d}=0.2$ that launch conical jets with an opening angle $\theta_{\rm j}=0.1$ and an efficiency $\eta_{1,2}=0.1$.
    At the interaction point $z\sim z_{\rm dis}$, we assume a moderate jet Lorentz factor $\Gamma_{\rm j}=1.5$.
    Finally, we take a viewing angle $i=\arccos \beta_{\rm j}$ for simplicity, since then the Doppler factor is $\Delta = \Gamma_{\rm j}$.
    At such close distances to the launching point, jets are expected to be magnetically dominated and pair-dominated.
    The minimum Lorentz factor of the non-thermal energy distribution of electrons is given by Eq. \ref{eq:gmin} and results in $\gamma'_{\rm min} \approx 4$.
    The maximum Lorentz factor is determined by the balance between acceleration and cooling and strongly depends on the acceleration efficiency (Eq. \ref{eq:gmax}).
    For this reason, we consider two values for this parameter: $\zeta_{\rm acc}=10^{-4}$, namely a low acceleration efficiency, and $\zeta_{\rm acc}=\beta_{\rm rec}=0.2$, namely maximum acceleration efficiency.
    For these cases, we obtain $\gamma'_{\rm max} \approx 10^4$ and $5\times 10^5$, respectively.
    These parameter values are summarised in Table \ref{tab:anysymbols}.

\subsection{Spectral Energy Distributions}
\label{sec:SED}

    Figure \ref{fig:SED_fiducial} shows the SED for the reference model including the emission from the CBD, the mini-discs, and the dual jet interacting region.
    We also show the emission from the counter-jets, which is a factor $\approx \Gamma_{\rm j}^8\approx25$ times smaller than the jet emission.
    The CBD and the mini-discs produce a thermal spectrum of comparable luminosity (for this separation), with peaks at UV frequencies of $\sim 2 \times 10^{15}~{\rm Hz}$ and $\sim 3 \times 10^{16}~{\rm Hz}$, respectively.
    The jet interacting region emits a non-thermal spectrum consisting of two superposed broad bumps due to synchrotron emission (red line in Fig. \ref{fig:SED_fiducial}) and SSC (purple line). 
    The synchrotron spectrum is divided into three regions.
    Below the SSA frequency $\nu_{\rm ssa}$ \citep[see, e.g.,][]{Ghisellini2013}, the emission is self-absorbed and the spectrum behaves as $F_\nu \propto \nu^{5/2}$.
    For frequencies higher than $\nu_{\rm ssa}$ and lower than the peak frequency of emission for the most energetic electrons, $\nu_{\rm max} \approx \nu_{\rm syn}(\gamma_{\rm max}') = \Gamma_{\rm j} (3/2)\gamma_{\rm max}'^2 \nu_{\rm L}$, the synchrotron spectrum is a power-law of index $\alpha \approx p/2 = 1.05$; here, $\nu_{\rm L}= eB'/2\pi m_{\rm e}c$ is the Larmor frequency.
    If the maximum Lorentz factor is given by Eq. \ref{eq:gmax}, then $\nu_{\rm max}$ only depends on the acceleration efficiency and the Lorentz factor: $\nu_{\rm max} \sim 9 \times10^{18} (\Gamma_{\rm j}/1.5) \zeta_{{\rm acc},-4}~{\rm Hz}$. This gives
    values of $\lesssim 10^{19}~{\rm Hz}$ and $\approx 10^{22}~{\rm Hz}$ for the cases with low and high acceleration efficiency, respectively.
    Finally, above $\nu_{\rm max}$, the synchrotron spectrum decreases exponentially.
    
    The SSC spectrum ranges from $\sim \gamma_{\rm min}'^2 \nu_{\rm ssa} \sim 2 \times 10^{15}$ Hz to the Klein--Nishina limit for the most energetic electrons: $E_{\rm max}^{\rm (ssc)}\sim \Gamma_{\rm j} \gamma'_{\rm max} m_{\rm e}c^2$ ($\lesssim 10~ {\rm GeV}$ for $\zeta_{\rm acc}=10^{-4}$ and $\approx 300$ GeV for $\zeta_{\rm acc}=0.2$).
    Nevertheless, the total spectrum is strongly affected by $\gamma \gamma$ absorption above $\nu_{\gamma \gamma} \approx 2-3 \times 10^{21}$ Hz, where the luminosity is attenuated by a factor $\tau_{\gamma \gamma}^{-1} \propto \nu^{-b}$, with $b\approx 1$.
    For $\nu_{\rm max} < \nu < \nu_{\gamma \gamma}$, the SSC spectrum is the result of the up-scattering of the power-law synchrotron photons and then shares the same spectral index $\alpha$.
    In the scenario with efficient acceleration, however, we have $\nu_{\gamma \gamma} 
    \sim \nu_{\max}$ and the SSC spectrum is completely absorbed.

\subsubsection{Dependance on the model parameters}

\begin{figure*}
        \includegraphics[width=0.44\linewidth]{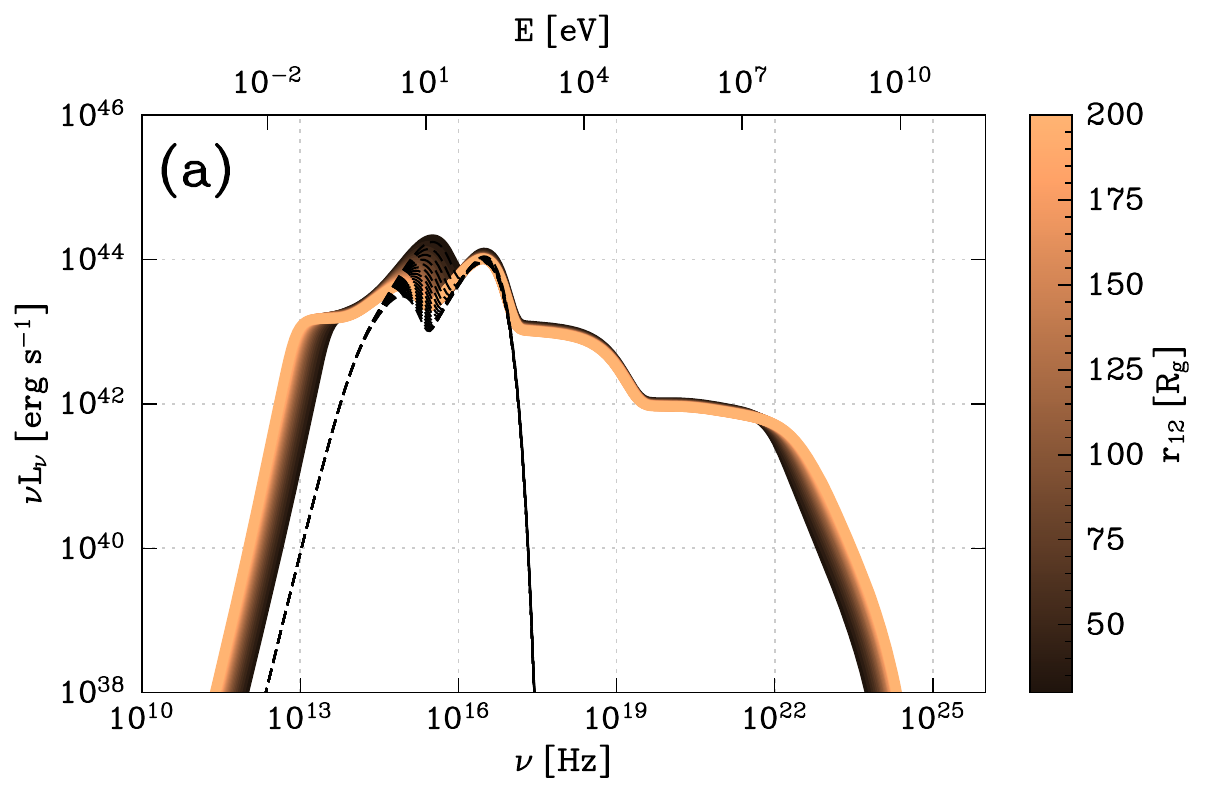}
        \includegraphics[width=0.44\linewidth]{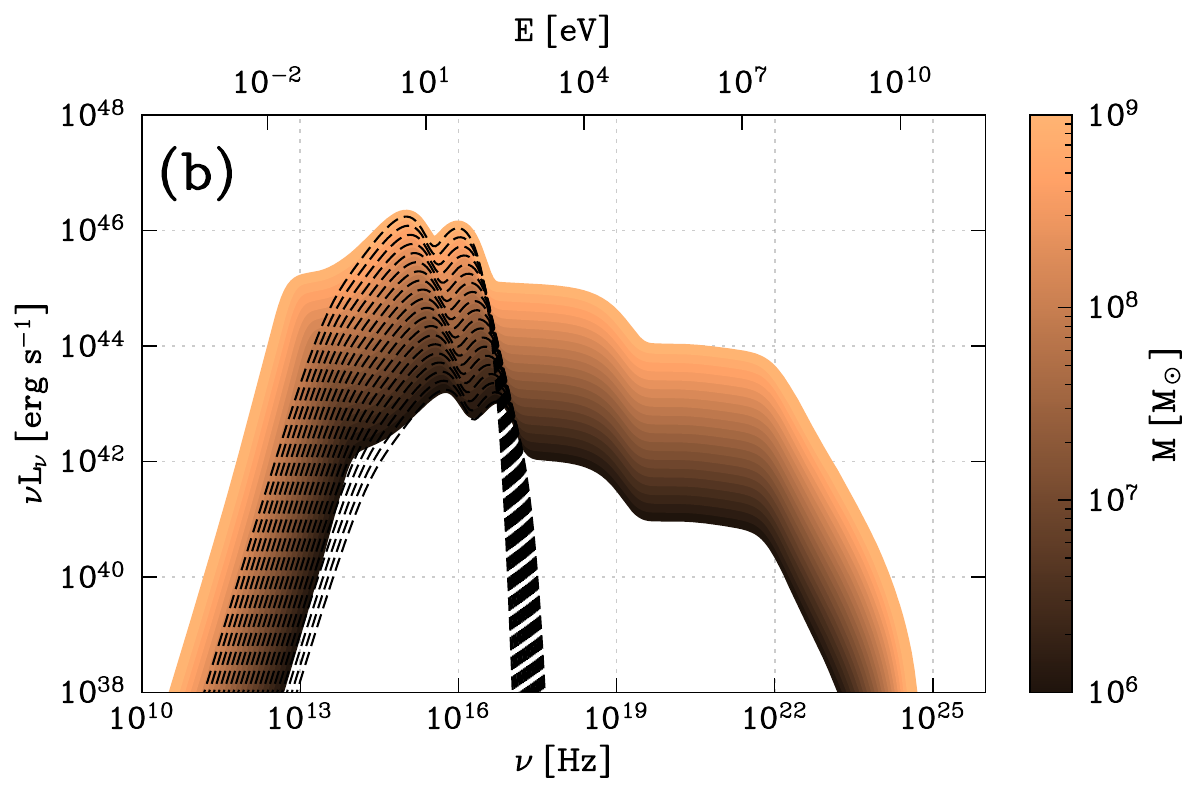}
        \includegraphics[width=0.44\linewidth]{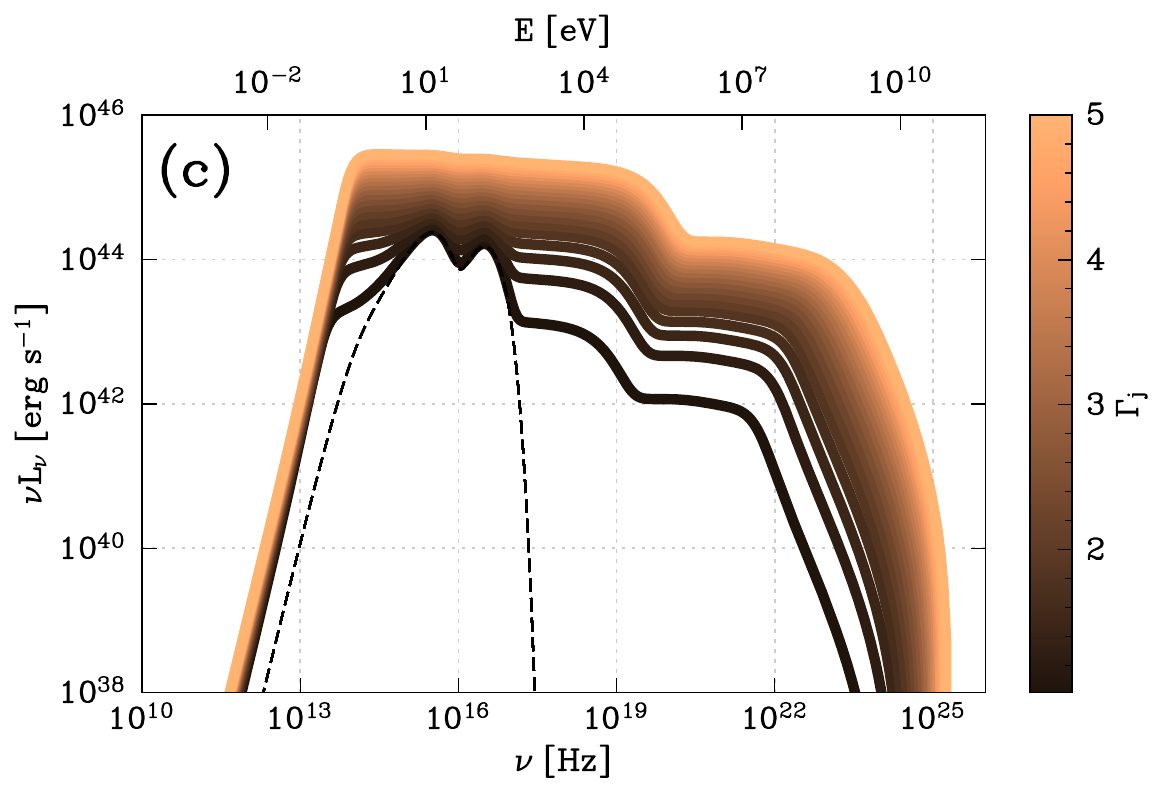}
        \includegraphics[width=0.44\linewidth]{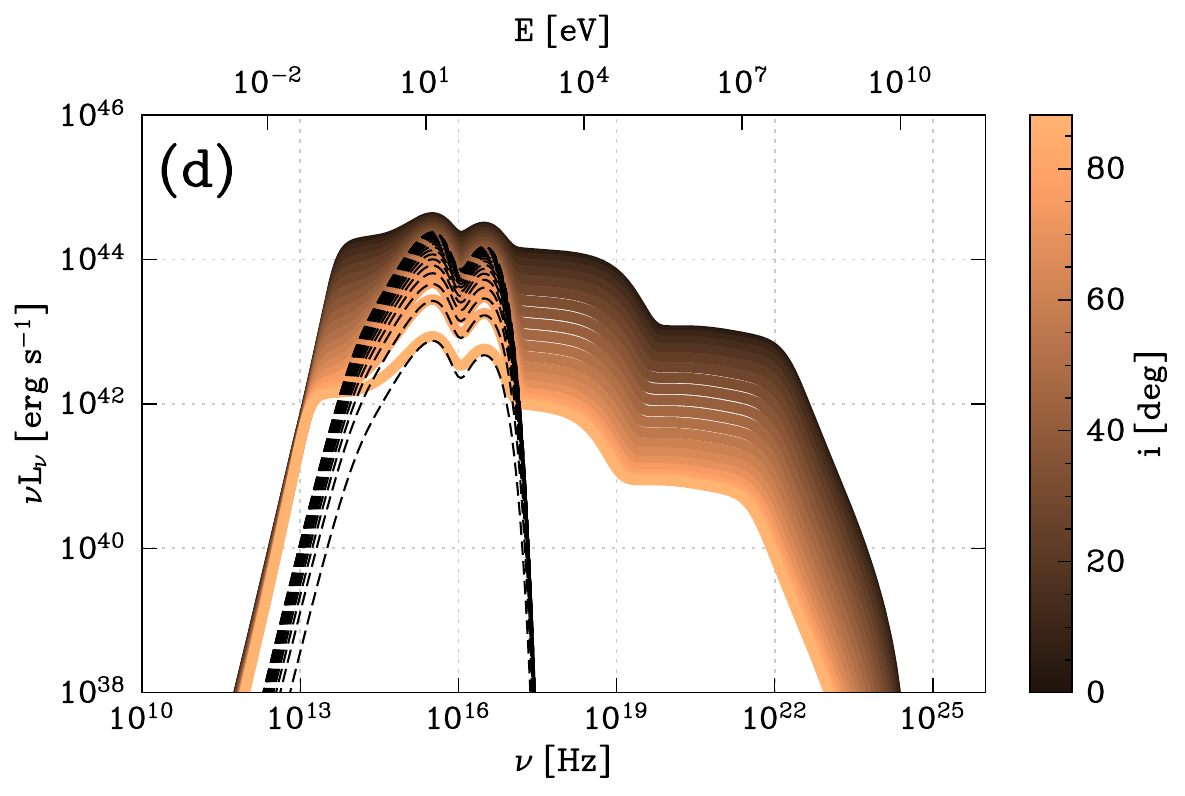}
        \caption{Spectral energy distribution of an SMBHB with dual jet emission for four scenarios where, in each of them we vary: a) the black hole separation $r_{12}$, b) the total mass of the system $M$, c) the Lorentz factor $\Gamma_{\rm j}$, and d) the viewing angle $i$. The remaining parameters are fixed to their values in the reference model, except for $i$ in Panel c) where we fixed it to $i=0^\circ$.
        The dashed lines show the CBD+mini-disc emission whereas the solid lines show the total emission after absorption.}
        \label{fig:SED_par}
\end{figure*}

    We now explore how the SED varies under changes in the values of the model parameters.
    Figure \ref{fig:SED_par} shows SEDs for four scenarios where in each of them one parameter is varied in a given range: a) the black hole separation $r_{12}$ between $30$ and $200~R_{\rm g}$; b) the total mass of the system between $10^6$ and $10^9~M_\odot$; c) the jets' Lorentz factor $\Gamma_{\rm j}$ between $1$ and $5$; and d) the viewing angle $i$ between $0$ and $90^\circ$.
    In each case, the remaining parameters are fixed to the reference values shown in Table \ref{tab:anysymbols}, except for the scenario ``c'', where we take $i=0^\circ$ so that the Doppler factor $\mathcal{D}$ monotonically increases with $\Gamma_{\rm j}$.

    Panel (a) shows that the SED changes very little with the black hole separation.
    This is because the particle energy distribution is dominated by synchrotron cooling and thus the synchrotron power is of the order of the injected electron power: $L_{\rm sy} \sim L_{\rm e}' \propto B'^2 l^2 \sim {\rm const.}$ (see Sec. \ref{sec:particle}), since $B' \propto r_{12}^{-1}$ (Eq. \ref{eq:B}) and $l \sim r_{12}$.
    The spectrum however is broader for larger separations, where the emission region is larger (less compact) and then $\nu_{\rm ssa}$ is lower, and $\nu_{\gamma \gamma}$ is larger.
    This implies that during the evolution of the orbit, as the orbit shrinks, the emission shows a secular flux decrease in both the low-energy and the high-energy ends of the spectrum (IR and $\sim 10~$MeV, respectively, for the reference model).
    
    The CBD luminosity does change with separation since its inner edge depends on it, $r_{\rm edge}\sim 2r_{12}$, and then its luminosity is $\propto M \dot{M}_{\rm cbd} r_{\rm edge}^{-1} \sim \dot{m} (r_{12}/R_{\rm g})^{-1}$.   
    The mini-disc spectrum is also expected to vary with $r_{12}$, since for higher separations, the ratio $r_{\rm t}/r_{\rm isco}$ would be larger and mini-discs would be more similar to normal disks; this, in turn, would imply a higher value of $f_{\rm d}$. However, this is not reflected in our model as we have kept this parameter fixed.

    Panel (b) shows how the SED varies with the system's total mass.
    Here, the dual jet luminosity scales linearly with the total mass as $\nu L_\nu \propto L_{\rm e}' \propto B'^2 l^2 \propto M_7$, whereas $\nu_{\rm ssa} \propto M_7^{-0.36}$.
    As expected, the frequency of the thermal peak of the discs' spectrum is proportional to the maximum temperature, which is $T_{\rm disc} \propto M_7^{-0.25}$, whereas the luminosity is  $\propto r_{12}^2T_{\rm disc}^4 \sim M_7$.
    
    Finally, in Panels (c) and (d), we show the dependence of the SED on the Doppler factor, by changing either the jets' Lorentz factor at the dissipation height $z_{\rm dis}$ or the viewing angle.
    For the first case, we fixed the line-of-sight viewing angle to $0^\circ$, so that the Doppler factor (and hence the luminosity enhancement) increases monotonically with the Lorentz factor: $\mathcal{D} = [\Gamma_{\rm j}(1-\beta_{\rm j})]^{-1} = \Gamma_{\rm j} (1+\beta_{\rm j}) $.
    For $\Gamma_{\rm j} \gtrsim 2.5$, the dual jet spectrum overcomes the discs', and it reaches a maximum of $L\sim 5\times 10^{45}~{\rm erg~s}^{-1}$ for $\Gamma_{\rm j}\approx 5$.
    On the other hand, for a fixed value of $\Gamma_{\rm j}$, a change in $i$ affects not only the Doppler factor but also the discs' emission, because the greater the viewing angle, the smaller the discs' effective surface: $L_{\rm disc} \propto \cos i$.
    This maintains an approximately constant ratio between dual jets' and discs' luminosity.
    
\subsubsection{Magnetisation}
\label{sec:SED_sigma}

\begin{figure}
    \includegraphics[width=0.9\linewidth]{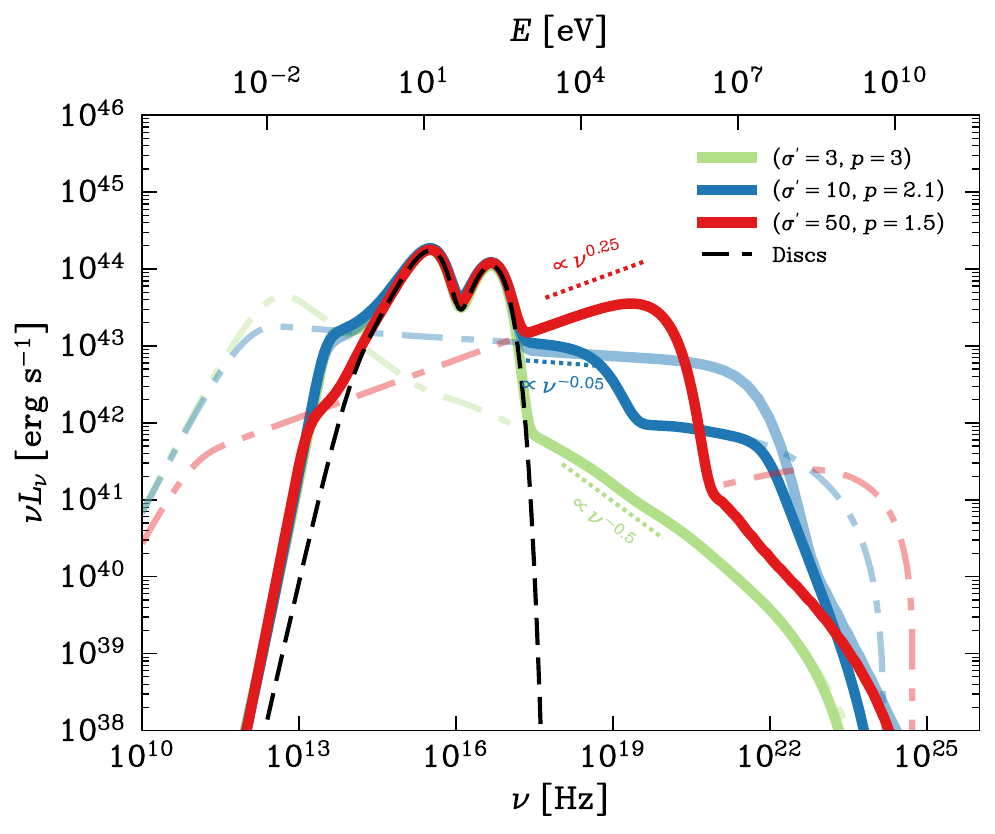}
    \caption{Spectral energy distribution for an accreting SMBHB with dual jet emission for different values of the jet magnetisation (dot dashed lines). For $\sigma'=3$, we take $p=3$; for $\sigma'=10$, we take $p=2.1$; and for $\sigma'=50$, we take $p=1.5$.
    For the reference model $(\sigma'=10,~p=2.1)$, we show the two models for low and high-acceleration efficiency.
    We also show the spectral indices for the power-law part of the spectrum: $\nu F_\nu \propto \nu^{-\frac{s-1}{2}+1}$.
    Dashed black lines show the emission from the CBD+mini-discs and solid lines show the total emission.}
    \label{fig:SED_sigma}
\end{figure}

    We consider separately the SED dependence with the magnetisation $\sigma'$, since it changes qualitatively the shape of the non-thermal spectrum.
    The magnetisation determines the maximum (if $\sigma' \gg 1$) and minimum particle energy (Eqs. \ref{eq:gmin} and \ref{eq:gmax_hard}) and the spectral index of the distribution, $p$, while it also influences, though weakly, the magnetic field and the total reconnected power.
    Following the results by \citet{Petropoulou_etal2016}, we explore three scenarios of low, medium, and high magnetisation of the jets.
    These three scenarios have different values for the pair $(\sigma', p)$, which are $(3, 3)$, $(10, 2.1)$, and $(50, 1.5)$, respectively.
     The other parameters are fixed to the values of the reference model.
    
    Figure \ref{fig:SED_sigma} shows the SED for these three scenarios.
    The dot-dashed dark line shows the discs' emission which is independent of magnetisation, while the dashed coloured lines show the dual jet non-thermal emission without absorption.
    Solid lines correspond to the total emission for the three scenarios, corrected by SSA and $\gamma \gamma$ absorption.
    The SSA frequency is approximately constant for the three scenarios, which then show the same spectrum at low energies.
    The SEDs are a power-law of index $-p/2+1$ between soft X-rays (where the mini-disc spectrum decays exponentially) and the minimum between $\nu_{\rm max}$, determined by $\gamma_{\rm max}'$, and $\nu_{\gamma \gamma}$.
    For $\sigma'=50$, the maximum energy of the particle distribution is limited by $\sigma'$ and reaches an intermediate value between the two limits set by the synchrotron losses in the reference model (this can be lower for a low acceleration efficiency).
    The spectrum is $\nu L_\nu \propto \nu^{0.25}$ up to an energy $\epsilon_{\rm max} \approx 330~{\rm keV}$, where the synchrotron spectrum decays exponentially, whereas the SSC spectrum is fully absorbed; above this energy, the SED decays as $\nu L_\nu^{\rm (ssc)} / \tau_{\gamma \gamma} \propto \nu^{-\frac{p}{2}+1}/\nu \sim \nu^{-0.75}$.
    For $\sigma'=3$, the SED goes as $\propto \nu^{-0.5}$ until $\nu_{\gamma \gamma}$, where the spectral index softens to $\approx 1.5$.

    We conclude that the higher the magnetisation, the higher the non-thermal signal at energies above the mini-discs thermal spectrum end.

\subsection{Emission variability}
\label{sec:variab}

    If the energy injection in the collision region is modulated in time, either by large-scale motions of the jets or by accretion rate variability at BH scales, the non-thermal radiation from the interaction can inherit this variability.
    As we discuss below, the outcome is different if the jets are aligned ($\theta_{\rm s} \approx 0$) or misaligned ($\theta_{\rm s} > \theta_{\rm j}$), where $\theta_{\rm s}$ is the inclination angle of the black hole spin.

\subsubsection{Misaligned jets}

    In a gas-rich environment, BHs with arbitrary spin directions will interact with the accretion flow in a process that will tend to align the spins with the angular momentum of the disc \citep{BP75, Liska_etal2019, Liska_etal2021}.
    In the case of SMBHBs, there are still large uncertainties on whether spin directions align completely or not before the merger \citep{Bogdanovic_etal2007, Dotti_etal2010, LodatoGerosa2013, Gerosa_etal2015, Gerosa_etal2020, Nealon_etal2022}. 
    If the BHs maintain misaligned spins during the inspiralling phase, jet launching can be strongly affected.
    Simulations of tilted accretion discs onto single BHs have found that both the inner disc angular momentum and the jets launched by the BZ mechanism tend to align with the BH spin \citep{McKinney_etal2013}.
    Such an alignment may be weak (e.g. in standard thick discs) or very strong (e.g. in thin magnetically arrested discs) \citep{Liska_etal2018}.
    At larger distances from the engine, however, the jet bends and aligns with the angular momentum of the outer disc due to pressure forces exerted by the corona and winds.
    In the case of binary accretion, such an alignment of the jets with the angular momentum of the CBD may happen much further away from the black holes than in the single BH case.
    This is so because the relatively small size of the mini-discs ($r_{\rm t} \sim 0.3 r_{12}$) and the truncation of the CBD far from the horizons (at $r_{\rm edge} \sim 2 r_{12}$) will significantly decrease the pressure forces exerted to the outflow in the cavity region.

    Given the above considerations, an SMBHB with misaligned spins may launch misaligned jets, which then, due to the orbital motion, would interact with each other only during a fraction of the orbit.
    The process occurs in the following way: as the two jets get close enough carried by the orbital motion, each of them will start to feel the magnetic pressure of the other jet.
    As the pressure increases, the magnetic flux will start to pile up on the edges, until the jets merge, bounce back, disrupt each other, or tunnel across  \citep{linton2001reconnection}.
    Although a robust understanding of the dynamics and radiation of colliding jets requires detailed numerical simulations, we give here an order-of-magnitude estimation for the timescales associated with the envisioned interaction.
    
    Let us approximate the jets as two flux tubes approaching with a speed $\sim 2v_{\rm bin} \sim \Omega_{\rm bin} r_{12}$ towards each other.
    The contact surface between the jets will have a length scale of the order of the jet cylindrical radius $R \sim \theta_{\rm j} z_{\rm dis}$.
    After the flux tubes touch, each jet will start to compress and get deformed due to the magnetic pressure of the other; we quantify this deformation by a non-dimensional parameter $\epsilon$, such that $\epsilon R$ is the length scale of the deformation.
    Under these conditions, a current sheet will form along the contact surface provided the compression speed is of the order of the reconnection speed:  $\dot{\epsilon} R \sim v_{\rm rec} \lesssim v_{\rm A}$, where we have assumed relativistic reconnection \citep{lyutikov2003dynamics,zweibel1994magnetic}.
    The maximal deformation corresponds to $\epsilon \sim 1$, which sets a characteristic timescale $\Delta t_{\rm int} \sim \epsilon / \dot{\epsilon} \sim \dot{\epsilon}^{-1} \sim R / v_{\rm rec}$, and gives an estimation for the duration of the interaction.
    Considering that $z_{\rm dis} \sim r_{12}/(2\theta_{\rm s})$ (see Sec. \ref{sec:jets}), we can rewrite this timescale as
\begin{equation}
    \Delta t_{\rm int} \approx 7.3 \times 10^3 M_7 \left( \frac{r_{12}}{30R_{\rm g}} \right) \left( \frac{\beta_{\rm rec}}{0.2} \right)^{-1} \left( \frac{\theta_{\rm j}}{\theta_{\rm s}} \right)~{\rm s}.
    \label{eq:delta_tint}
\end{equation}

\begin{figure}
    \centering
    \includegraphics[width=0.999\linewidth]{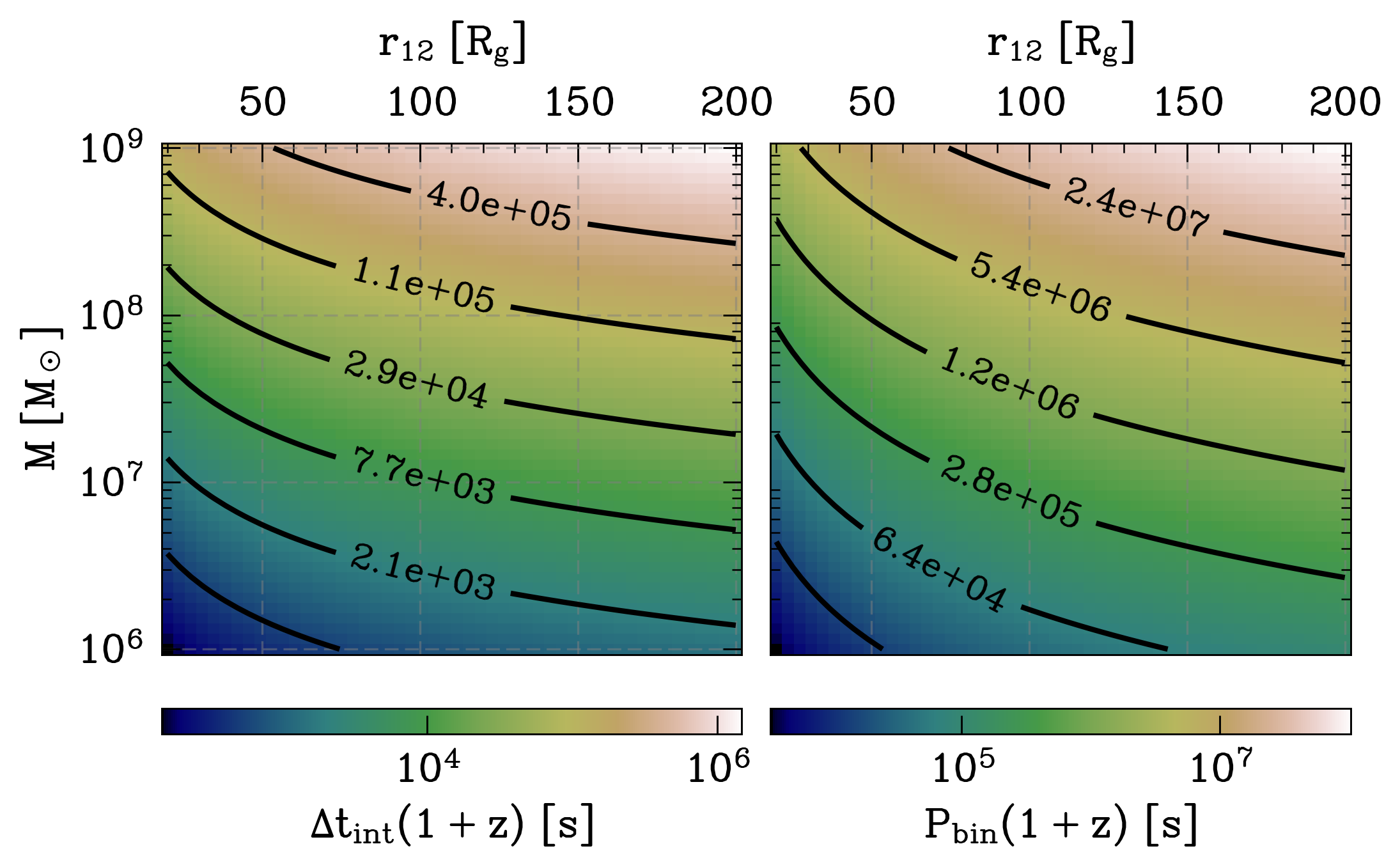}
    \caption{Contour plot of the characteristic timescales in the observer frame for the flare duration (left panel, $\Delta t_{\rm int}$) and flare period (right panel, equal to $P_{\rm bin}$) as a function of the binary separation (in geometrised units) and the total mass of the system. The redshift is $z=0.2$.}
    \label{fig:timescales}
\end{figure}

    The interaction would be followed by a period of quiescence until the jets encounter each other again after one orbit.
    In summary, these jet encounter events might give rise to non-thermal flares with a duration $\Delta t_{\rm int}$ (Eq. \ref{eq:delta_tint}) that repeat periodically after $\sim P_{\rm bin}$ (Eq. \ref{eq:Pbin}).
    Figure \ref{fig:timescales} shows these two timescales as a function of the geometrised binary separation, $r_{12}/R_{\rm g}$, and the total mass of the system, $M$; we have fixed $v_{\rm rec}=0.2c$, $\theta_{\rm s}=0.2$, $\theta_{\rm j}=0.1$, and considered a redshift $z=0.2$.
    The duration of the flares goes from $\sim $ minutes to $\sim $ months, while the period of repeating ranges from $\sim$ hours to $\sim$ years, increasing with both black hole separation and total mass.

\subsubsection{Aligned jets}

    As we discussed in the previous sections, aligned dual jets would interact at a height $z_{\rm dis} \sim r_{12}/(2\theta_{\rm j})$ due to the non-zero jet opening angle.
    Because the interaction is persistent rather than temporary as in the misaligned case, the potential variabilities associated with the non-thermal emission will be mainly mediated by the intrinsic variability in the Poynting flux.
    This, in turn, is mediated by the accretion rate in the mini-discs (see Eq. \ref{eq:Lj}), which means that MHD variabilities induced by the binary accretion can be translated to the dual jet non-thermal emission.

    The CBD emits a non-periodic thermal spectrum that only varies secularly as the binary orbit shrinks due to GW emission.
    In contrast, the mini-discs' emission may oscillate periodically as $\sim \cos (\omega t)$ due to at least two effects: the modulation of the accretion rate, $f_i(t)$, for example due to the presence of a `lump' in the CBD (see \citealt{Bowen2018, Bowen2019, Combi2021b}), $\omega \sim \Omega_{\rm bin} - \Omega_{\rm lump} \sim 0.7 \Omega_{\rm bin}$, and the relativistic beaming (and de-beaming) imprinted by the fast orbital motion of the black holes\footnote{At the fiducial separation of $r_{12}\approx 30 R_{\rm g}$, the orbital velocity of the black holes is $v_{\rm orb} \sim 0.18 c$, and the luminosity ratio between the maximum and minimum of the cycle can be as large as $\sim 4$ for an edge-on view.}, $\omega \sim \Omega_{\rm bin}$ \citep{dorazio2015nature}.
    The accretion modulation, if present, would be translated to the jet power and thus to the non-thermal emission, which would occur with a delay $\gtrsim z_{\rm dis} / c $ to the mini-disc emission.

    Finally, additional potential modulation in the non-thermal emission can arise due to the jet precession, which in SMBHBs may occur due to inertial forces produced by the orbital movement, spin precession due to the spin-orbit or spin-spin coupling, or mini-disc precession.
    All these effects would induce modulations with periods longer than the orbital period.

\section{Discussion}
\label{sec:discussion}

   A fraction of the AGNs in the universe produces powerful relativistic jets that can be radio-loud on kpc scales and also radiate high-energy emission from X-rays to $\gamma-$rays \citep{ulrich1997variability, fossati1998unifying}.
   Jetted AGN spectra typically show a double-humped structure formed by a low-energy hump, likely due to synchrotrons from electrons in the jets, and a high-energy hump from hard X-rays to $\gamma-$rays, c.f. \citep{rani2019multi, bloom1996}.
   There are still uncertainties within models to explain the origin of the high-energy component, which can be leptonic (SSC or external inverse Compton with soft photons from the broad line region or the dusty torus) or hadronic (pp or p$\gamma$ emission), e.g. \citep{petropoulou2015constraints, celotti2008power}. 
   The main mechanisms responsible for the energy dissipation in AGN jets are magnetic reconnection and shocks.
Because these sources are bright both in radio and $\gamma$ rays ($>10$ GeV), the dissipation region should not be too compact to avoid self-absorption, and is likely located outside the broad line region, at $\sim$ pc-scales from the BH, or otherwise it would show absorption features \citep{costamante2018origin}.
   The variability observed at high energies, on the contrary, suggests that this component may come from more compact regions \citep{aleksic2011magic}, and it can be of the order of minutes to hours.
   This is thought to arise mainly due to internal processes in the emitting region and is strongy affected by relativistic contraction: $\delta t_{\rm obs} \ll \delta t'$.

    The emission from dual jet interaction we envisioned here presents some key differences to that of single black hole jets.
   First, the characteristic variability timescales associated with dual jet emission would be comparable to the binary orbital period, ranging from hours to years depending on the separation and total mass of the system (see Figure \ref{fig:timescales}).
   If the jets are misaligned, the nonthermal emission will only be present when the jets encounter each other once per orbit, while the CBD and mini-disk emission is persistent.
Though difficult to unambiguously determine its origin, detecting such flaring radiation would be a promising smoking gun to identify dual jet emission from SMBHBs.

   Second, dual jet interaction provides a natural mechanism for a strong magnetic dissipation of the jet power at very close distances to the black holes.
   The emission region is then very compact and magnetised, and the resultant spectrum is self-absorbed both at low and high energies; we predict little emission above $\sim (10-100) ~ {\rm MeV}$.
   Hence, efforts to look for variability in dual jets should be favoured from future sub-GeV observatories such as AMEGO-X \citep{Caputo_etal2022}, e-ASTROGRAM \citep{DeAngelis_etal2017}, and GRAMS \citep{Aramaki_etal2020}.
We also notice that the emission from a dual jet interaction can be largely correlated to the UV/soft X-ray bands associated with the mini-disk emission.
In the case of aligned spins, the dominant modulations in the dual jet emission would likely be linked to modulations in the accretion rate onto the binary cavity, but its nonthermal spectrum would cover a much broader range than the thermal emission from the accretion disks.

   We have centered our analysis on SMBHBs at relatively small separations, $r_{12} \lesssim 100 r_{\rm g}$.
Dual jet interaction from larger separation binaries might also result in strong non-thermal emission, but the scenario may differ from the magnetically-dominated case we envisioned here.
The interaction will occur much farther from the engine where the jets have a lower magnetisation and a larger inertia.
Hence, the dissipation will likely be dominated by other processes such as shocks triggered by the dual jet collision.
The resultant interacting region would be less compact decreasing the effect of internal absorption at low and high energies.

    On the other hand, we have focused on equal-mass binaries, which will likely be common at close separations \footnote{In an unequal-mass accreting binary, the secondary receives most of the mass and grows more rapidly than the primary, leading the system to unity mass-ratios if the accretion rate is high enough during the whole evolution \citep{noble2021}}.
At larger separations, unequal-mass binaries would be more common and the dual jet structure would be different from our scenario, with one jet much more powerful than the other.
The presence of double jets, in this case, might translate into periodic perturbations in the dominant jet's emission.
This could be similar to the aligned scenario considered in our work, though with a much lower perturbation amplitude in the emission.

\section{Summary and conclusions}
\label{sec:conclusions}

    We have developed a model to calculate the emission produced by the interaction of dual jets launched by SMBHB in the GW-dominated regime.
    We have shown that the interaction region between two magnetised jets is prone to form large-scale reconnection layers where magnetic energy is released.
    Part of this energy is transferred to the particles from the jets, which are accelerated and form a non-thermal distribution.

    Our main conclusions and findings can be summarised as follows:
    \begin{itemize}
    
        \item  Magnetic reconnection accelerates particles in the collision region.
        These particles cool by synchrotron radiation and SSC.
        Because the jet interaction region is compact and highly magnetised, the emission is internally absorbed at low frequencies ($ \lesssim 10^{13}$ Hz) due to SSA and at high-energies ($ \gtrsim 10 $ MeV) due to $\gamma \gamma$ absorption.
        The non-thermal SED shape strongly depends on the magnetisation of the jets and peaks at energies $\lesssim $ MeV if $\sigma > 10$ and $p<2$ (see Fig. \ref{fig:SED_sigma}).
        
        \item  The luminosity and SED from dual jet interactions is not greatly affected by the separation of the binary, as long as the jets are magnetically dominated when they interact ($z_{\rm dis} < 10^3$ R$_{\rm g}$).

        \item  If the jets are misaligned with the angular momentum of the binary, the interaction can occur only during some parts of the orbit, giving rise to periodic non-thermal flares (see Fig. \ref{fig:timescales} for characteristic timescales for the duration and period of these events).
        If the jets are aligned, the non-thermal emission could be modulated similarly to the mini-disc emission, but with a time delay.
        
        \item  Interacting dual jets suffer strong magnetic dissipation at close distances to the BHs, and thus the emission is more absorbed than in single AGNs both at low and high energies.
        Variability, when present, would also be different from most AGN/Blazar sources since it would be linked to the orbital motion of the BHs.
        
    \end{itemize}

We conclude by emphasising that much further numerical work will be needed to fully understand the properties of dual jet interactions, both in the microphysical realm, e. g. through plasma simulations of relativistic colliding flux-tubes, and in the macrophysics realm, e.g through large-scale GRMHD simulations of jet production in SMBHBs.

\section*{Acknowledgements}

%The Acknowledgements section is not numbered. Here you can thank helpful
%colleagues, acknowledge funding agencies, telescopes and facilities used etc.

We thank the anonymous referee for useful comments that significantly helped to improve the paper.
We also thank Jens Mahlmann, Luca Comisso, Bart Ripperda, Xinyu Li, Huan Yang, Will East and Nils Simonsen for fruitful discussions about plasmas and magnetic reconnection physics.
We have extensively used the open-source tools provided by \texttt{Matplotlib} \citep{matplotlib}, \texttt{NumPy} \citep{numpy}, 
 and \texttt{SciPy} \citep{scipy}.
E. M. G. and L. C received support from CONICET (Argentina's National Research Council) fellowship program and the Rochester Institute of Technology's Center for Computational Relativity and Gravitation.
EG acknowledges funding from the National Science Foundation under grant No AST-2108467 and from an Institute for Gravitation and the Cosmos fellowship.
LC is a CITA National fellow and acknowledges the support of the Natural Sciences and Engineering Research Council of Canada (NSERC), funding reference DIS-2022-568580.
Research at Perimeter Institute is supported in part by the Government of Canada through the Department of Innovation, Science and Economic Development Canada and by the Province of Ontario through the Ministry of Colleges and Universities.
Additionally, M.C. gratefully acknowledges support from NSF awards NSF AST-2009330, OAC-2031744 and PHY-1806596, PHY-2110352 and NASA TCAN grant (NNH17ZDA001N).
Computational resources were provided by the TACC's Frontera supercomputer allocation No. PHY-20010 and AST-20021.
Additional resources were provided by the RIT's BlueSky and Green Pairie and Lagoon Clusters 
acquired with NSF grants PHY-2018420, PHY-0722703, PHY-1229173 and PHY-1726215.
GER acknowledges financial support from the State Agency for Research of the Spanish Ministry of Science and Innovation under grant
PID2022-136828NB-C41/AEI/10.13039/501100011033/, and by “ERDF A way of making Europe”, by the European Union, and through the "Unit of Excellence Mar\'ia de Maeztu 2020-2023" award to the Institute of Cosmos Sciences (CEX2019-000918-M). Additional support came from PIP 0554 (CONICET).

%%%%%%%%%%%%%%%%%%%%%%%%%%%%%%%%%%%%%%%%%%%%%%%%%%
\section*{Data Availability}

The data underlying this article will be shared on reasonable request to the corresponding author.
 
%The inclusion of a Data Availability Statement is a requirement for articles published in MNRAS. Data Availability Statements provide a standardised format for readers to understand the availability of data underlying the research results described in the article. The statement may refer to original data generated in the course of the study or to third-party data analysed in the article. The statement should describe and provide means of access, where possible, by linking to the data or providing the required accession numbers for the relevant databases or DOIs.



\bibliographystyle{mnras}
\bibliography{biblio, bhm_references} % if your bibtex file is called example.bib

%\begin{thebibliography}{99}
%\end{thebibliography}

%%%%%%%%%%%%%%%%%%%%%%%%%%%%%%%%%%%%%%%%%%%%%%%%%%

\end{document}